
\input phyzzx
\PHYSREV
\hoffset=0.3in
\voffset=-1pt
\baselineskip = 14pt \lineskiplimit = 1pt
\frontpagetrue
\rightline {Cincinnati preprint, July, UCTP-103-1995}
\medskip
\titlestyle{\seventeenrm Supersymmetric Wilson Loops,
Superstring-like Observables, and the Natural Coupling of
Superstrings
To Supersymmetric Gauge Theories.}
\vskip.5in
\medskip
\centerline {\caps M. Awada\footnote+{\rm E-Mail address:
moustafa@physunc.phy.uc.edu} and F.Mansouri\footnote*{\rm E-Mail
address:
Mansouri@UCBEH.SAN.UC.EDU}}
\centerline {Physics Department}
\centerline {\it University of Cincinnati,Cincinnati, OH-45221}
\bigskip
\centerline {\bf Abstract}
\bigskip

We obtain an explicit expression for the supersymmetric Wilson loop
in
terms of chiral superfields and supercurrents in superspace.  The
result
turns out to be different from what one would expects from the
simple
replacement of Lie algebraic valued connection in the exponent with
the
corresponding super-Lie algebraic one.  In the abelian
supersymmetric
gauge theories, generalizing the super-particle coupling
represented by
the exponent of the supersymmetric Wilson loop, we show there exist
a
unique dimensionless coupling of the superstring to abelian
supersymmetric
gauge theories that respects all known symmetries. The coupling is
expressed
in terms of chiral currents in superspace.  The natural
superstring coupling gives rise to a new observable that is
"stringy" in
nature and has no analogue in non-supersymmetric gauge theories.

\eject

Wilson loop observables [1,2] have found a variety of applications
in the
study of gauge theories, ranging from phenomenology to topological
field theories [3].  This is because they are gauge invariant
observables characterized by a dimensionless coupling constant.  In
abelian gauge theories they can be interpreted quantum mechanically
as the amplitude of the
creation and annihilation of $e^{-}e^{+}$ resulting from the unique
coupling between the Maxwell's field and the point particle with
dimensionless
coupling constant.  In non-abelian gauge theories, this
interpretation can be extended formally, in the sense that the
particle-Yang-Mills coupling can only be realized by the Wilson
loop
$$ W_{R}(C) = Tr Pe^{ie\oint_{C} A}\eqno{(1)}$$
in a given representation R of the Lie algebra.  In equation (1)
the symbol
Tr stands for the trace, P for path ordering, and C is a closed
path. At short distance scales perturbative calculations shows that
the
Wilson loop produces the perimeter law indicating short range
finite forces between quarks as well as the renormalization of the
Coulomb potential between two heavy static quarks.  These results
indicate
that quarks can propagate freely at short distances.  On the other
hand in
lattice gauge theories strong coupling expansion shows that the
Wilson loop
produces the Area law suggesting confinement.
Unfortunately there is no known analytical method of extracting the
linear
potential or the area law from the Wilson loop in the continuum.
This raises the question whether the area law or the string can
arise
naturally through some dimensionless coupling to gauge theories.
It can be easily seen that there is no consistent dimensionless
coupling
with at most two derivatives between strings and gauge theories
except
for the parity violating expression
$$  ie\int d\sigma F^*\  .\eqno{(2)}$$
In (2) $d\sigma$ is the surface measure, and $F^*$ is
the dual of the gauge field strength.  However, there do exist
higher
derivative dimensionful couplings between strings and gauge fields
[4].

In the last two years, there have been significant developments in
N=1 and N=2 supersymmetric gauge theories [5].  In particular a
mechanism
for confinement was provided by the condensation of monopoles.
In finite supersymmetric gauge theories, such as N=4 super
Yang-Mills theory,
and a class of N=2 super Yang-Mills theories coupled to N=2 matter
the
corresponding beta functions vanish and the issues of confinement
and
asymptotic freedom become unclear.   Therefore, it is of interest
to
explore whether the supersymmetric extension of the Wilson loop or
the
existence of new supersymmetric observables might shed some light
on the
above issues.

The aim of this letter is two fold: One is to generalize the usual
notion of a Wilson loop to the supersymmetric case. The other is to

show that, unlike the non-supersymmetric gauge theories, there is
a new
observable in supersymmetric gauge theories whose current is
chiral and has support only on a two surface with a boundary.  Its
generalization to an arbitrary surface gives rise to a unique
and natural coupling between the superstring and the
supersymmetric gauge theories, which is characterized by a
dimensionless
coupling constant.  This allows us to construct a new "stringy"
observable,
in addition to the supersymmetric Wilson loop, that respects all
the
symmetries of the theory.

We begin with the supersymmetric Wilson loop.  For clarity of
presentation,
we give the details for the supersymmetric Maxwell's theory.  The
generalization to the supersymmetric non-abelian gauge theories
will be
presented in a second article. Throughout this paper we will use
the superspace two component notation.

In the expression (1) for the Wilson loop, the notion of a
connection plays
a dominant role.  So it might appear at first sight that to obtain
a
suitable extension of (1) for the supersymmetric gauge theories, we
only
need to replace the Lie algebra valued connection with a
superalgbera
valued connection and replace the trace instruction with the
supertrace one.  Although such a prescription works in some special
cases for locally
supersymmetric gauge theories, such as Chern-Simons supergravity
theories
in 2+1 dimensions [6], it does not work for globally supersymmetric
gauge theories.  The underlying reason is the structure of the
superpartners of the gauge potential in the two cases.  The
superpartner
of the former carries a vector index, while that of the latter does
not.

To look for another point of departure, we return to the exponent
of the
Wilson loop which has the structure of the interaction between the
point particle and the abelian gauge field represented uniquely by
a
dimensionless coupling constant $e$:
$$S_{int} = ie\oint_{C} d\lambda A(x(\lambda)).{\dot x}(\lambda)
\  .\eqno{(3)}$$
Stoke's theorem implies that the interaction lagrangian can be
rewritten in terms of the field strength of the gauge field on
a closed two surface $\Sigma$ whose boundary is the loop $C$:
$$S_{int} = {ie\over 2}\int_{\Sigma(C)} d^2\xi \epsilon^{ab}F_{ab}
\eqno{(4)}$$
where  $\xi^{a}; a=1,2$ is the coordinate of the two world sheet,
and [4]
$$F_{ab} := v_{a}^{\mu} v_{b}^{\nu}F_{\nu\mu}=
\partial_{a}A_{b}-\partial_{b}A_{a}\eqno{(5)}$$
where the field $A_{a}$ is defined to be the projection of the
gauge field
$A_{\mu};\mu=0..3$ along the two surface:
$$ A_{a} := v_{a}^{\mu}A_{\mu}~~;~~v_{a}^{\mu} =
\partial_{a}x^{\mu}(\xi)\  .\eqno{(6)}$$
It is important to remark that the Wilson loop is a topological
entity.
Furthermore the fact that there is no difference between the two
definitions
(3) and (4) of the Wilson loop can be interpreted to mean that the
extra
surface coordinate, say, $\xi^{1}$, is not dynamical and
can be integrated over.
\bigskip
{\bf The Supersymmetric Wilson loop}
\medskip
The important advantage of expressing the SUSY Wilson loop in terms
of superfields is that it will be manifestly supersymmetric and
gauge invariant.
We start by generalizing the interaction (4) to that of a
superparticle
coupled to a supersymmetric abelian gauge field theory.
The definition in (6) can be generalized in superspace using a
super one-form $\Gamma_{A}$ and the standard supersymmetric
notation:
$$F_{ab} = v_{a}^{A}v_{b}^{B}F_{BA}\eqno{(7)}$$
where
$$F_{AB} = D_{[A}\Gamma_{B)} - T_{AB}^{C}\Gamma_{C}\eqno{(8)}$$
and T is the torsion supertensor. The
bases $v$ have to be constructed from the super-coordinate
$z^{M}=(x^{\mu},
\theta^{m},\theta^{{\dot m}})$ in such a way that they are
supersymmetric
invariant.  We define
$$v_{a}^{A} = E_{M}^{A}\partial_{a}z^{M}\eqno{(9 a)}$$
where $ E_{M}^{A}$ is a veilbein and obtain the following
components:
$$v_{a}^{\alpha{\dot \alpha}} =\partial_{a}x^{\alpha{\dot
\alpha}}(\xi)
-{i\over 2}(\theta^{\alpha}(\xi)\partial_{a}\theta^{{\dot
\alpha}}(\xi)
+\theta^{{\dot \alpha}}(\xi)\partial_{a}\theta^{\alpha}(\xi))$$
$$v_{a}^{\alpha} =\partial_{a}\theta^{\alpha}(\xi)\eqno{(9 b)}$$
$$v_{a}^{{\dot \alpha}} =\partial_{a}\theta^{{\dot \alpha}}(\xi).$$
The $v$'s are invariant under the global space-time supersymmetry
transformation rules defined
$$\delta x^{\alpha{\dot \alpha}}(\xi) = {i\over
2}(\epsilon^{\alpha}
\theta^{{\dot \alpha}}(\xi) +\epsilon^{{\dot
\alpha}}\theta^{\alpha}(\xi))$$
$$\delta \theta^{\alpha}(\xi) = \epsilon^{\alpha}\eqno{(9 c)}$$
$$\delta \theta^{{\dot \alpha}}(\xi) = \epsilon^{{\dot \alpha}}.$$
The requirement that the coordinates $\theta$ satisfy the Majorana
condition
demands that $\epsilon$ be a Majorana.  From the $v$'s we will
construct the
following supersymmetric and gauge invariant tensors:
$$ C_{ab}^{{\dot \alpha}{\dot \beta}}={i\over 2}
v_{a \beta}^{({\dot \alpha}}v_{b}^{{\dot \beta})\beta}~~;
 C_{ab}^{\alpha\beta}={i\over 2}
v_{a {\dot \beta}}^{(\alpha}v_{b}^{\beta){\dot \beta}}\eqno{(10
a)}$$
$$C_{ab}^{\alpha} :=v_{a}^{\alpha{\dot \alpha}}v_{b {\dot \alpha}}
\eqno{(10 b)}$$
$$C_{ab}^{{\dot \alpha}} :=v_{a}^{\alpha{\dot \alpha}}v_{b
\alpha}.$$
The vector components of the C's are antisymmetric in a and b by
construction ( the parenthesis denotes symmetrization with unit
weight factor) and obey the following properties:
$$C_{ab}^{{\dot \alpha}{\dot \beta}} ={1\over 2}\epsilon_{ab}
C^{{\dot \alpha}{\dot \beta}}~~;~~C^{{\dot \alpha}{\dot \beta}}
=\epsilon^{ab}C_{ab}^{{\dot \alpha}{\dot \beta}}$$
$$ C_{ab}^{\alpha\beta} ={1\over 2}\epsilon_{ab}C^{\alpha\beta}~~;
{}~~C^{\alpha\beta} =\epsilon^{ab}C_{ab}^{\alpha\beta}\  .\eqno{(10
c)}$$
Note, however, the spinor components of C do not obey any
particular symmetry.  This crucial property allows us to use these
components to construct a
new type of interaction with supersymmetric gauge theories that are

inherently "stringy" as we will show below.  For later use we also
note that
$$ C_{[ab]}^{\alpha} = {1\over 2}\epsilon_{ab}C^{\alpha}~~;
{}~~C^{\alpha} =\epsilon^{ab}C_{ab}^{\alpha}$$
$$C_{[ab]}^{{\dot \alpha}} ={1\over 2}\epsilon_{ab}
C^{{\dot \alpha}}~~;~~C^{{\dot \alpha}}
=\epsilon^{ab}C_{ab}^{{\dot \alpha}}\eqno{(10 d)}$$
where the square bracket indicates antisymmetrization.
To get the correct content of fields of super Maxwell`s theory
or super Yang-Mills theory one has to impose representation
preserving constraints whose purpose is to project the superfield
formulation onto chiral superspace where the supersymmetry
representation is irreduciable.  The required constraints are [7]
( and references therein):
$$F_{\alpha{\dot \alpha}} = F_{\alpha\beta} =F_{{\dot \alpha}
{\dot \beta}}=0\  .\eqno{(11 a)}$$
Together with the Bianchi identities of $F_{AB}$, one can determine
the
remaining components of F in terms of a chiral superfield
$(W_{\alpha},W_{{\dot \alpha}})$.  For the abelian supergauge
theory one obtains:
$$F_{\alpha,\beta{\dot \beta}} = \epsilon_{\alpha\beta}W_{{\dot
\beta}}~~;
{}~~F_{{\dot \alpha},\beta{\dot \beta}} =
\epsilon_{{\dot \alpha}{\dot \beta}}W_{\beta}$$
$$F_{\alpha{\dot \alpha},\beta{\dot \beta}} ={-i\over 2}
(\epsilon_{\alpha\beta}D_{({\dot \alpha}}W_{{\dot \beta})} +
\epsilon_{{\dot \alpha}{\dot \beta}}D_{(\alpha}W_{\beta )})\eqno
{(11 b)}$$
where the $W$'s satisfy the chirality conditions:
$$D_{\alpha}W_{{\dot \beta}} = D_{{\dot \alpha}}W_{\beta } =
0\eqno{(12 a)}$$
and
$$D^{{\dot \alpha}}W_{{\dot \alpha}} = D^{\alpha}W_{\alpha}\
.\eqno{(12 b)}$$
In turn the $W$'s are determined in terms of an unconstrained
vector
superfield $V$:
$$W_{\alpha} = {-i\over 2}{\bar D}^2D_{\alpha}V~~;~~ W_{{\dot
\alpha}} =
{i\over 2}D^2D_{{\dot \alpha}}V\  .\eqno{(13)}$$
They are invariant under the gauge transformation
$$\delta V = i({\bar \Lambda} - \Lambda)\eqno{(14)}$$
where $\Lambda$ (${\bar \Lambda}$) is a chiral (anti-chiral)
parameter superfield.
The component expansion of $V$ and $W_{\alpha}$ in the Wess-Zumino
gauge are
respectively,
$$V = (0,0,0,0, A_{\alpha{\dot \alpha}},\psi_{\alpha},\psi_{{\dot
\alpha}}, D)
\eqno{(15 a)}$$
$$W_{\alpha} = \psi_{\alpha} - \theta^{\beta}f_{\alpha\beta}
-i\theta_{\alpha}D
+{i\over 2}\theta^{2}\partial_{\alpha{\dot \alpha}}\psi^{{\dot
\alpha}}
\eqno{(15 b)}$$
where $\psi$ is the superpartner of the gauge field $A_{\alpha{\dot
\alpha}}$,
$f_{\alpha\beta} = {1\over 2}\partial_{(\alpha{\dot
\alpha}}A_{\beta)}^{{\dot
\alpha}}$ is the Maxwell's field strength and D is an
auxiliary field.  Another important property of the $W_{\alpha}$
($W_{{\dot \alpha}}$) is that it is invariant under the chiral
(anti-chiral) supersymmetry transformations of the component
fields:
$$\delta_{\epsilon^{\alpha}}W_{\alpha} =
\delta_{\epsilon^{{\dot\alpha}}}W_{{\dot \alpha}} = 0\  .\eqno{(16
a)}$$
$$\delta A^{\alpha{\dot \alpha}} = i(\epsilon^{\alpha}\psi^{\dot
\alpha}
+\epsilon^{{\dot\alpha}}\psi^{\alpha})$$
$$\delta \psi_{\alpha} = \epsilon^{\beta}f_{\alpha\beta}+
i\epsilon_{\alpha}D\eqno{(16 b)}$$
$$\delta D ={1\over 2}\partial_{\alpha{\dot\alpha}}
(\epsilon^{\alpha}\psi^{\dot \alpha}
-\epsilon^{{\dot\alpha}}\psi^{\alpha})$$

Equipped with (11) and using the tensor definitions (10) we deduce
from (8) that the interaction action between the superparticle
and the supersymmetric gauge theory is given by
$$ S_{int}^{(1)} ={1\over 2}ie\int_{\Sigma(C)} d^2\xi
\epsilon^{ab}F_{ab}
\  .\eqno{(17 a)}$$
But now the field strength $F_{ab}$ is given by
$$F_{ab} = \epsilon_{ab}({1\over 2}C^{\alpha\beta}(\xi)D_{\alpha}
W_{\beta}(x(\xi),\theta(\xi)) + C^{\alpha}(\xi)W_{\alpha}(x(\xi),
\theta(\xi)) + h.c)\eqno{(17 b)}$$
where h.c denotes hermitian conjugate.  We will shortly
show that (17) can be neatly rewritten in terms of chiral currents
in
superspace.  The abelian supesymmetric Wilson loop is now given by
$$ W(C) = e^{S_{int}^{(1)}}\eqno{(18)}$$
\eject
\bigskip
{\bf A Superstring-like Observable}
\medskip
Looking at the structure of the spinor components of C in (10) one
immediately recognizes the existence of a new two form which is not
topological but a dynamical interaction that has support only
on the surface:
$$S_{int.}^{(2)} = \kappa\int_{\Sigma(C)}d^2\xi \sqrt{-h} h^{ab}
C_{ab}^{\alpha}(\xi)W_{\alpha}(x(\xi),\theta(\xi)) +
h.c\eqno{(19)}$$
where $h^{ab}$ is the metric on the two surface and h is its
determinant.
We know of no way to construct such a term in the absence of
supersymmetry.
Moreover, unlike the expression (18), this interaction cannot be
reduced to
an expression on the line.
It is supersymmetric, gauge, and reparemetrization invariant and
characterized by a new dimensionless  coupling constant $\kappa$
which is different from $e$ classically. Thus we can define a new
super-gauge
invariant observable
$$\Psi(C) = e^{S_{int.}^{(2)}}\eqno{(20)}$$
whose correlation function might be of potential interest for
strongly
coupled super QED.
\bigskip
{\bf  Chiral Currents}
\medskip
The combination of the Wilson loop and the superstring-like
observable
can be totally expressed in terms of chiral currents on the
surface.
Define
$$ J^{\alpha}(z) = q^{\alpha}(z) - D_{\alpha}q^{\alpha\beta}(z)
\eqno{(21 a)}$$
where
$$q^{\alpha}(z) = \int_{\Sigma(C)}d^2\xi (eC^{\alpha}(\xi) +
\kappa\sqrt{-h} h^{ab}C_{ab}^{\alpha}(\xi))\delta^{6}(z-z(\xi))
\eqno{(21 b)}$$
$$q^{\alpha\beta}(z) ={1\over 2}\int_{\Sigma(C)}d^2\xi
(eC^{\alpha\beta}(\xi))\delta^{6}(z-z(\xi))\eqno{(21 c)}$$
and $\delta^{6}(z-z(\xi)) =\delta^{4}(x-x(\xi))(\theta^2
-\theta^2(\xi))$
is the chiral superspace delta function.  The full interacting
actions
(17) and (19) are therefore given by:
$$S_{int} = S_{int}^{(1)}+S_{int}^{(2)} =
\int d^6z (J^{\alpha}W_{\alpha} + h.c)\  .\eqno{(22)}$$
which is manifestly supersymmetric and gauge invariant.
We could have discovered the interacting lagrangians (17) and (19)
by
postulating (22) on basis of dimensionality, gauge, and
supersymmetric invariance.  The dimension of $W$ is $[W]= -{3\over
2}$ in units of
length. Since $[d^6 z]=+3$ we deduce that $[J] =-{3\over 2}$.
If $J$ is a chiral current with support on the two surface
$$ J^{\alpha}(z) =\int_{\Sigma(C)}d^2\xi
j^{\alpha}(\xi)\delta^{6}(z-z(\xi))\eqno{(23)}$$
then $j^{\alpha}(\xi)$ is a two dimensional supersymmetric and
reparametrization invariant spinor of dimension $-{1\over 2}$.
The only candidate objects to construct such a $j^{\alpha}(\xi)$
are from the tensors $C$ in equation (10).
In particular the spinor components have the correct dimension of
$-{1\over 2}$.  The most general structure of $j^{\alpha}(\xi)$ is
$$j^{\alpha}(\xi) = \kappa\sqrt{-h} h^{ab}C_{ab}^{\alpha}(\xi)
+{\tilde e}C^{\alpha}(\xi) +{e\over
2}C^{\alpha\beta}(\xi)D_{\alpha}
\eqno{(24)}$$
where the super-derivative is understood to act on the chiral delta
function in (23).  The fact that the surface has a boundary,
together
with the Bianchi identity (12b) and current conservation implies
that
$e ={\tilde e}$, however if the surface is closed with no boundary,
then
$e$ and ${\tilde e}$ are different in general.  Of course in the
absence
of supersymmetry $C_{ab}^{\alpha}=0$ and the current (24) reduces
to that
of the Wilson loop.  Having obtained the supersymmetric Wilson loop
and a
superstring-like observable, we note that (19) can be viewed as
a new type of interaction that corresponds to the interaction of a
superstring with the abelian supersymmetric gauge theory if the
chiral current (23) is defined over an arbitrary surface without a
boundary.
In this context, equation (22) represents
the interaction of both the superparticle and the superstring with
an
abelian supersymmetric gauge theory characterized by dimensionless
coupling constants.  Integrating out the gauge field from the
expression
(19) will lead to an
effective superstring theory as a consequence of short distance
renormalization.  A natural question which arises is whether such
a
superstring theory is one of the known varieties or is a new one.
We will have more to say about this elsewhere [8].
Another important question is whether such a superstring theory or
a superstring-like observable can exist in a $D+1$ dimensional
Minkowski
world of the form $M_{D}$x$R$.  This existence depends crucially
on whether or not the closed surface $\Sigma$ can be embedded
in $M_{D}$.  For $D=3$ the answer is certainly negative in general.

However, for simply connected 3-mainfolds such as the three sphere,

the embedding exists.  Furthermore as shown and discussed in [9],
there is a whole class of non-simply connected manifolds in which
any loop $C$ embedded in them can be thought of as the boundary of
a closed surface $\Sigma$ in $M_{3}$.

To summarize, the main results of this work may be stated, a
posteriori,
in the following way.  We have obtained a general gauge and
supersymmetric
invariant expressions for the interaction of the superparticle and
the
superstring with a supersymmetric abelian gauge field.  When
restricted
to a surface with a boundary, these expressions lead to the
supersymmetric
Wilson loop and a superstring-like observable which has no counter
part in
non-supersymmetric gauge theories.

This work was supported, in part, by the department of energy under
the
contract number DOE-FG02-84ER40153.

{\bf References}

\item {[1]} K.G.Wilson, Phys.Rev. D10 (1974) 2455.
\item {[2]}  A.M. Polyakov, Phys. Lett. B59 (1975) 82;
F. Wegner, J.Math. Phy. 12 (1971) 2259
\item {[3]} E. Witten, Comm. Math. Phys. 117, 353 (1988).
\item {[4]} L.N.Chang, and F. Mansouridvi2ps - x 10 susy |lpr -Plw4
proceeding of the John Hopkins
workshop, ed. G.Domokos and S.Kovesi Domokos, John Hopkins Univ.
(1974).
\item {[5]} N. Sieberg and E. Witten, Nucl. Phys. B426 19 (1994).
\item {[6]} K. Koehler, F. Mansouri, C. Vaz, and L. Witten, Nucl.
Phys. B348
,373, (1990).
\item {[7]} J. Wess and J. Bagger, Introduction to supersymmetry,
Princeton
University Press 1983.
\item {[8]} M. Awada and F. Mansouri, in preperation.
\item {[9]} M.Awada, Comm. Math. Phys. 129;329 (1990).

\end